\documentstyle[12pt,epsf]{article}

\def\beq{\begin{equation}}
\def\eeq{\end{equation}}
\def\bea{\begin{eqnarray}}
\def\eea{\end{eqnarray}}
\def\bq{\begin{quote}}
\def\eq{\end{quote}}

\def\bq{\begin{quote}}
\def\eq{\end{quote}}

\def\bq{\begin{quote}}
\def\eq{\end{quote}}

\def \lsim{\mathrel{\vcenter
     {\hbox{$<$}\nointerlineskip\hbox{$\sim$}}}}

\def\gappeq{\mathrel{\rlap {\raise.5ex\hbox{$>$}}
{\lower.5ex\hbox{$\sim$}}}}
\def\lappeq{\mathrel{\rlap{\raise.5ex\hbox{$<$}}
{\lower.5ex\hbox{$\sim$}}}}

\def\bbz{fa Z \kern-8.9pt Z}
\def\ZZ{\hbox{\it Z\hskip -4.pt Z}}

\begin{document}

\baselineskip 24pt

\newcommand{\sheptitle}
{Does LEP prefer the NMSSM?}

\newcommand{\shepauthor}
{M. Bastero-Gil$^{1}$, C. Hugonie$^{2}$, S. F. King${^1}$, D. P. Roy$^{3}$,
S. Vempati$^{4}$}

\newcommand{\shepaddress}
{$^{1}$Department of Physics and Astronomy, University of Southampton,\\ 
Southampton, SO17 1BJ, U.K.\\ 
$^{2}$Rutherford Appleton Laboratory,\\ 
Chilton, Didcot, Oxon, OX11 0QX, U.K.\\
$^{3}$Theoretical Physics Department, Tata Institute of Fundamental Research,\\
Homi Bhabha Road, Bombay 400 005, India\\
$^{4}$Theory Group, Physical Research Laboratory,\\
Navarangpura, Ahmedabad 380 009, India.}

\newcommand{\shepabstract}
{We study the naturalness of electroweak symmetry breaking and baryogenesis in
the next-to-minimal supersymmetric standard model (NMSSM). Our study is
motivated by the recent LEP bounds on the Higgs boson mass which severely
constrains the low $\tan \beta$ region of the minimal supersymmetric standard
model (MSSM). We show that the low $\tan \beta$ region of the NMSSM is clearly
favoured over the MSSM with regard to the physical Higgs boson mass,
fine-tuning, and electroweak baryogenesis.}

\begin{titlepage}
\begin{flushright}
TIFR/TH/00-23 \\
SHEP/00-07 \\
RAL-TR/2000-026 \\
hep-ph/0006198\\
\end{flushright}
\begin{center}
{\large{\bf \sheptitle}}
\bigskip \\ \shepauthor \\ \mbox{} \\ {\it \shepaddress} \\ \vspace{.5in}
{\bf Abstract} \bigskip \end{center} \setcounter{page}{0}
\shepabstract
\end{titlepage}

\section{Introduction}

The latest LEP bound on the Standard Model like Higgs boson mass $m_h$ is in
the region of $m_h>108$ GeV once the limits from all the experiments are
combined \cite{one}. Such a bound on the Higgs mass can be interpreted as the
lightest scalar Higgs mass bound of $m_h > 105$ GeV in the 
low $\tan\beta (\leq
4)$ region of the minimal supersymmetric standard model (MSSM). Such a large
Higgs mass implies large fine-tuning in the low $\tan \beta$ region 
\cite{two},
and leads to the smallest values of $\tan \beta$ being excluded. This is
disturbing since electroweak baryogenesis in the MSSM relies on low $\tan
\beta$ and light Higgs and stop masses \cite{three} \footnote{When SUSY phases
are included the parameter space for electroweak baryogenesis is somewhat
increased, however \cite{kane}.}.

It is well known that in the next-to-minimal supersymmetric standard model
(NMSSM) \cite{four} the lightest Higgs boson can be heavier than in the MSSM
\cite{five}. It is also well known that the parameter space for electroweak
baryogenesis in the NMSSM is much larger than in the MSSM, due to trilinear
contributions to the scalar potential at tree-level. 
What has not been realised
so clearly is that these two features go together with low
$\tan\beta$. 
This is
a timely observation, given the constraints that LEP places on the MSSM in the
low $\tan\beta$ region. Moreover, although fine-tuning has been well 
studied in
the MSSM it has not been systematically studied in the NMSSM. In this letter,
then, we compare the fine-tuning in the NMSSM and MSSM in the low $\tan\beta$
region and show that the NMSSM is much preferred. We then study the 
strength of
the electroweak phase transition (EWPT) in the NMSSM, and show that, unlike in
the MSSM, a sufficiently strong first order phase transition persists
over 
much
of the low $\tan \beta$ region.

Although the NMSSM provides a testable solution to the $\mu$ problem, by
replacing the superpotential term $\mu H_1H_2$ by $\lambda N H_1H_2$ (where
$H_i$ are Higgs doublets and $N$ is a Higgs singlet) it is often 
criticised for
leading to a conflict between cosmological domain walls and stability due to
supergravity tadpoles. It has recently been pointed out, however, that the
NMSSM remains a natural solution to the $\mu$ problem since both the 
stablility
and the cosmological domain wall problems may be eliminated by imposing a
$\ZZ_2$ R-symmetry on the non-renormalisable operators \cite{Greeks}. Thus the
NMSSM appears to be well motivated both theoretically and 
phenomenologically at
the present time. Indeed the solution of the $\mu$ problem is closely linked
to the two phenomenological features of NMSSM noted above in the sense that
all the three originate from the same term in the superpotential, as we shall
see below.

\section{Fine-Tuning in the NMSSM}

Although fine-tuning is not a well defined concept, the general notion of
fine-tuning is unavoidable since it is the existence of fine-tuning problem in
the standard model which provides the strongest motivation for low energy
supersymmetry, and the widespread belief that superpartners should be found
before or at the LHC. If one abandons the notion of fine-tuning then there is
no reason to expect superpartners at the LHC. Although a precise measure of
{\em absolute} fine-tuning is impossible, the idea of {\em relative}
fine-tuning can be helpful in selecting certain models and 
regions of parameter
space over others. It is useful to compare different models using a common
definition of fine-tuning \cite{six}
\beq
\Delta_a = abs\left(\frac{a}{M_Z^2}\frac{\partial M_Z^2}{\partial a}\right)
\label{one}
\eeq
where $a$ is an input parameter, and fine-tuning $\Delta^{max}$ is defined to
be the maximum of all the $\Delta_a$. It is worth pointing out that at low
values of $\tan \beta$ fine-tuning is worse in the MSSM for two separate
reasons. First, the tree level contribution to the Higgs mass squared upper
bound $M^2_Z \cos^2 2\beta$ goes down; so that one must rely more on radiative
corrections to meet the LEP bound, which demand a higher value of $M_3(0)$ as
observed in \cite{two}. Second, the top quark Yukawa coupling is 
larger for low
values of $\tan \beta$, so the Higgs mass gets driven more negative, resulting
in larger coefficients of the $M_3^2(0)$ term in Eq.~(\ref{one}), which again
increases fine-tuning. A quantitative study of fine-tuning reveals that it is
the experimental limit on the Higgs mass rather than the gluino mass, that
provides the most severe fine-tuning for low values of $\tan \beta$, as
discussed in the second reference in \cite{two}. Moreover the fine-tuning
required in the MSSM increases exponentially with the Higgs mass, since the
radiative corrections to the Higgs mass increase logarithmically with the stop
masses, and the stop masses increase in proportion to $M_3(0)$ which controls
the fine-tuning \cite{two}.

While the Higgs quartic coupling in the MSSM is related to the gauge
couplings, in the NMSSM it has an additional contribution from the
Yukawa coupling $\lambda$. Consequently the tree level mass limit of
the lightest Higgs scalar gets modified to 
\beq m_h^2 \leq M_Z^2
\left( \cos^2 2\beta + {2\lambda^2 \over g^2+g'^2} \sin^2 2\beta
\right) . \label{two} \eeq 
Note that the additional term in Eq.~(\ref{two}) is most
effective where its help is needed most, i.e. at low
$\tan\beta$. 
Assuming this Yukawa coupling to remain perturbative up
to the unification scale of $\sim 10^{16}$ GeV implies $\lambda \lsim
0.7$ at the electroweak scale \cite{five}, i.e. a contribution of
$\sim M^2_Z \sin^2 2\beta$ to the tree level mass limit. On the other
hand, the upper limit on the mass of the lightest CP even Higgs is not
necessarily physically relevant, since its coupling to the $Z$ boson
can be very small. Actually, this phenomenon can also appear in the
MSSM, if $\sin^2(\beta-\alpha)$ is small.  However, the CP odd Higgs
boson $A$ is then necessarily light ($m_A \sim m_h < M_Z$ at tree
level), and the process $Z \rightarrow h A$ can be used to cover this
region of the parameter space in the MSSM. In the NMSSM, a small gauge
boson coupling of the lightest CP even Higgs is usually related to a
large singlet component, in which case no (strongly coupled) light CP
odd Higgs boson is available. 
Hence, Higgs searches in the NMSSM have
possibly to rely on the search for the second lightest Higgs scalar,
which can be substantialy heavier than the limit given in
Eq.~(\ref{two}) \cite{seven}.   
In the limit that the singlet decouples, 
and is the heaviest scalar, 
the inequality in Eq.~(\ref{two}) may be  saturated
by the lightest Higgs boson. 
Alternatively if the singlet decouples, 
but is the lightest scalar, 
then the inequality in Eq.~(\ref{two}) may be saturated
by the second lightest scalar,
which is the physical Higgs boson. In fact this latter case approximately 
applies to the results we present later in the figures. 
In all cases it is clear that the NMSSM does not
require a large radiative correction to satisfy the Higgs mass limit
from LEP. It is this modification of the Higgs sector in the
NMSSM which is responsible for the opening up the low $\tan\beta$
region of parameter space in this model, by removing the connection
between the Higgs mass and exponential increases in fine-tuning
present in the MSSM. 

The NMSSM superpotential is defined as 
\beq W = \lambda N H_1 H_2 +
\frac{k}{3}N^3 + ... \label{three} 
\eeq 
where the elipsis stand for
quark and lepton Yukawa couplings, and the Higgs potential is 
\bea
V_{NMSSM} & = & m_1^2v_1^2 + m_2^2v_2^2 - 2m_3^2v_1v_2 +
\lambda^2v_1^2v_2^2 \nonumber \\ & + & \frac{1}{8}
(g'^2+g^2)(v_1^2-v_2^2)^2 + x^2(m_N^2 + \frac{2}{3} kxA_k+k^2x^2)
\label{four}
\eea
where $v_i = \langle H_i \rangle$, $x = \langle N \rangle$ and
\beq
m_1^2=m_{H_1}^2+\lambda^2x^2,~
m_2^2=m_{H_2}^2+\lambda^2x^2,~
m_3^2=-\lambda x(kx+A_\lambda).
\label{five}
\eeq
$A_{\lambda}, A_k$ are the trilinear soft parameters associated with the
$\lambda, k$ terms in the superpotential, which play a prominent role in
ensuring a strong 1st order EWPT, as we shall see in the next
section. One-loop
corrections to the effective potential, $\Delta V^{(1)}$, are taken into
account by redefining the scalar masses such that $m_i^2 \rightarrow m_i^2 +
\partial \Delta V^{(1)}/ \partial v_i^2$. The NMSSM minimisation 
conditions are
then
\bea
\frac{M_Z^2}{2} & = & \frac{m_1^2 - m_2^2 \tan^2\beta}{\tan^2\beta - 1}
\nonumber \\
\sin 2\beta & = & \frac{2m_3^2}{m_1^2 + m_2^2 + \lambda^2v^2} \label{six} \\
0 & = & \lambda x (2kx+A_{\lambda})\frac{v_1v_2}{x^2} + kx(2kx+A_k) +
(\lambda^2v^2+m_N^2) \nonumber
\eea
The results in Eq.~(\ref{six}) can then be used to derive a master 
formula for the
derivative of the $Z$ mass with respect to input parameters, 
from which one can
obtain the sensitivity parameter in Eq.~(\ref{one})
\bea
\frac{\partial M_Z^2}{\partial a} & = & \frac{2}{\tan^2\beta - 1} \left\{
\frac{\partial m_1^2}{\partial a} -\tan^2\beta \frac{\partial m_2^2}{\partial
a} -\frac{\tan\beta}{\cos 2\beta} \left( 1 + \frac{M_Z^2 - \lambda^2v^2}{m_1^2
+ m_2^2 + \lambda^2 v^2} \right) \right. \nonumber \\
& \times & \left. \left[ 2\frac{\partial m_3^2}{\partial a} - \sin 2\beta
\left( \frac{\partial m_1^2}{\partial a} +\frac{\partial m_2^2}{\partial a}
+\frac{\partial \lambda^2 v^2}{\partial a} \right) \right] \right\} .
\label{seven}
\eea
This result is very similar to the master formula obtained in the MSSM, to
which it reduces in the limit $\lambda \rightarrow 0$. Of course the partial
derivatives on the right-hand side will be quite different since they bring in
derivatives of $\lambda x$ rather than $\mu$, and $x$ is a function of all the
soft parameters (unlike $\mu$ in the MSSM which is independent of the soft
parameters.) Thus the implementation of the master formula in the 
NMSSM is more
involved than in the MSSM. Nevertheless, the variation of the VEVs 
follows from
the minimisation conditions of the potential, such that: 
\beq
\frac{\partial^2 V_{NMSSM}}{\partial a \partial v_i^2}=0 ,\quad
\frac{\partial^2 V_{NMSSM}}{\partial a \partial x^2}=0 . \label{eight}
\eeq
The above system of 3 coupled equations can then be solved for $\partial
v_i/\partial a$, and $\partial x/\partial a$. The other partial
derivatives we need
to evaluate the master formula Eq.~(\ref{seven}), such as $\partial \lambda/
\partial a$, $\partial m_{H_i}^2/\partial a \cdots$, are computed numerically
when running the renormalisation
group equations (RGEs). 

The NMSSM minimisation conditions are clearly analagous to those of the MSSM,
and they may be satisfied by an analagous proceedure. 
In both models the soft
parameters are chosen at the high energy scale. 
Once the (low energy) value of
$\tan\beta$ is selected, the top mass fixes the low energy top Yukawa
coupling\footnote{We include only QCD corrections when converting pole mass to
running mass at the $m_Z$ scale.}. In the NMSSM the additional low energy
values of $\lambda$ and $k$ are selected. Next, all the Yukawas are run up to
the high energy scale. Then all the soft masses are run down to low energies
and the value of $x$ (the analogue of $\mu$ in the MSSM) is fixed by the first
minimisation condition (up to an ambiguity in the signs), and the value of
$A_{\lambda}$ (the analogue of $B$ in the MSSM) is fixed by the second
minimisation condition. In the NMSSM it then only remains to satisfy the third
minimisation condition. In general this requires an iterative process since it
relies upon having chosen the correct values of $A_k$ and $m_N$ at high
energies (and both these parameters must be fixed before all the soft masses
can be run down). 
\footnote{In the limit that we neglect the
$\lambda$ and $k$ contributions to the right-hand sides of RGEs, the
RGEs for the soft masses $m_{Q}^2, m_U^2, m_{H_1}^2, m_{H_2}^2$
become identical to those in the MSSM, and that for
$A_{\lambda}$ becomes identical to that for the soft
parameter $B$ in the MSSM. 
Furthermore $m_N$, $A_k$ and $k$ do not run
in this limit. In this limit the first two minimisation conditions
are identical to those of the MSSM and the third can trivially
be satisfied by using it to fix $m_N$ at the end. Although we never
make this approximation, it serves to emphasise that main differences
between the NMSSM and the MSSM arise from the low energy Higgs
potential and scalar mass squared matrix not from RG running.}
The structure of the Higgs potential being more
complicated in
the NMSSM than in the MSSM, the minimisation conditions (\ref{six}) do not
guarantee that we sit at a local minimum rather than at a local maximum, in
which case the lightest squared mass eigenvalue is negative. The acceptable
regions of the low energy parameter space where the physical 
scalar squared masses
are positive were mapped out by Elliott et al in \cite{five}, and may be
straightforwardly be achieved by adjusting the high energy soft Higgs masses
$m_{H_1}(0)$ and $m_{H_2}(0)$. We therefore take a common value $m_0$ at
$M_{GUT}$ for the squark and slepton masses, but allow the input Higgs masses
to be different. Also, a positive squared Higgs mass spectrum is favoured when
$A_k(M_Z)$ is small, so we have adjusted the initial value of $A_k(0)$ such
that $A_k(M_Z)=0$.

\begin{figure}[t]
\epsfxsize=10cm
\epsfxsize=10cm
\hfil \epsfbox{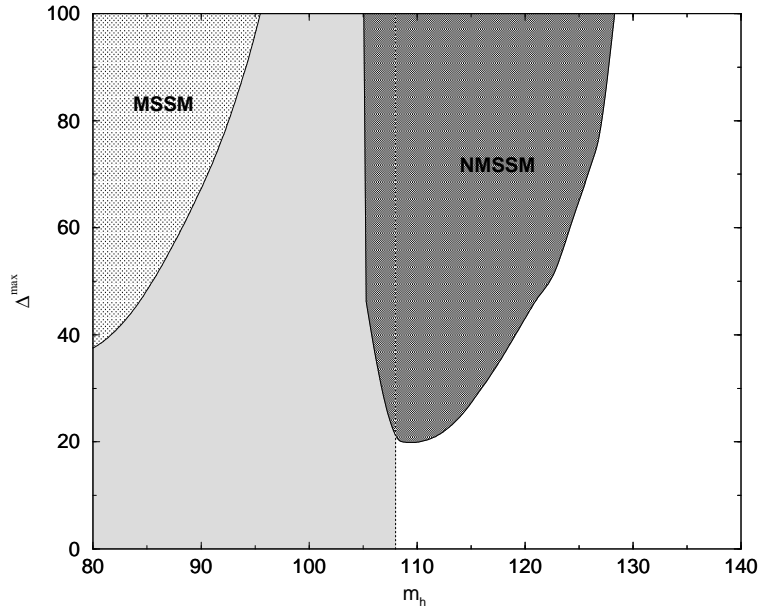} \hfil
\caption{{\footnotesize The maximum sensitivity parameter $\Delta^{max}$ as a
function of the lightest physical Higgs mass for $tan\beta=2$. We have fixed 
$m_0=100$ GeV, $A_t(0)=0$ GeV, $M_2(0)=M_1(0)=500$ GeV and $\mu <0$ 
($\lambda x <0$), while
varying $M_3(0)$, $m_{H_1}(0)$ and $m_{H_2}(0)$. The darkest grey region
correspond to the NMSSM, and the lighter grey region to the MSSM.
The thin dotted line at 108 GeV, and the lightest shaded region to the
left of it, represents the LEP excluded region on the standard
model Higgs boson mass.}} 
\label{tan2}
\end{figure}

\begin{figure}
\epsfxsize=10cm
\epsfxsize=10cm
\hfil \epsfbox{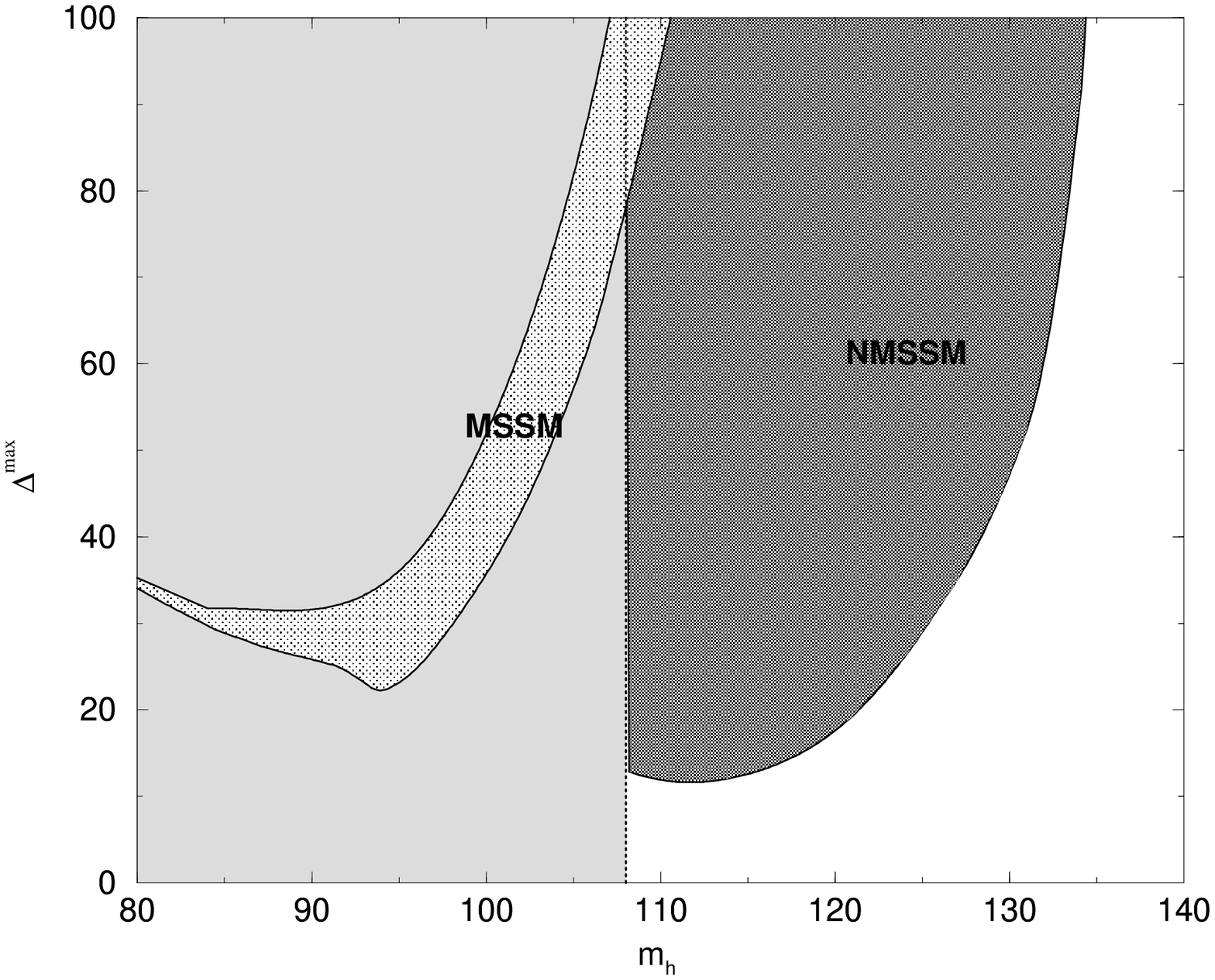} \hfil
\caption{{\footnotesize Same as in Fig. (1) for $\tan\beta=3$.}}
\label{tan3}
\end{figure}

\begin{figure}
\epsfxsize=10cm
\epsfxsize=10cm
\hfil \epsfbox{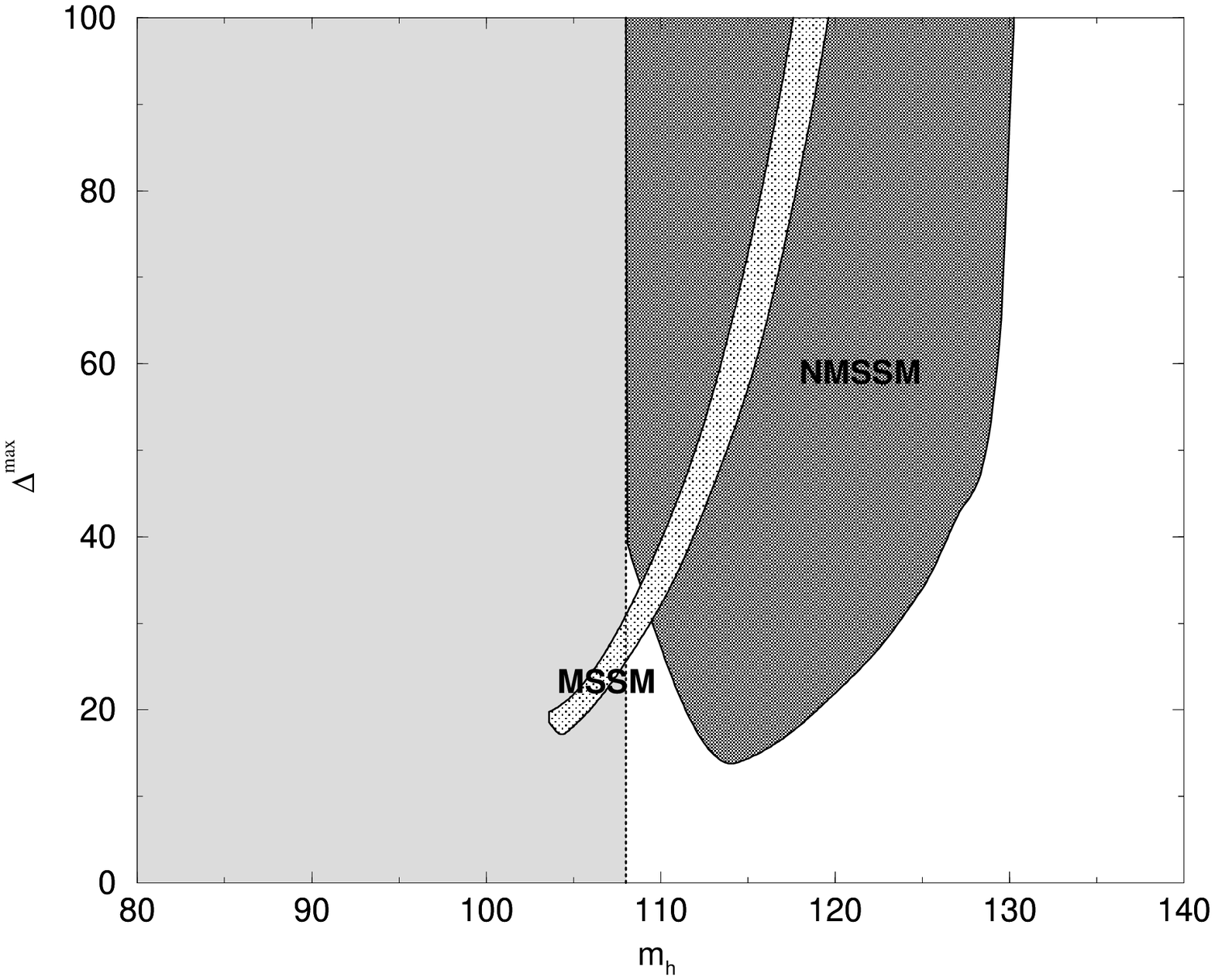} \hfil
\caption{{\footnotesize Same as in Fig. (1) for $\tan\beta=5$.}}
\label{tan5}
\end{figure}

In figs. \ref{tan2}-\ref{tan5} we plot the maximum sensitivity parameter
$\Delta^{max}$ as a function of the lightest physical Higgs mass for both the
NMSSM and MSSM, for the values of $\tan\beta=2,\,3,\,5$. 
In these plots we have taken
$m_0=100$ GeV, $A_t(0)=0$ GeV, and assumed high energy first and 
second gaugino massses given
by $M_2(0)=M_1(0)=500$ GeV. In our scans over parameter space we have
restricted ourselves to $100\, {\rm GeV}< M_3(0) < 600\, {\rm GeV}$, 
$0 < m_{H_1}(0) < 1\,
{\rm TeV}$, and $\mu<0$ ($\lambda x <0$). In the MSSM, for each pair of points
($M_3(0),\;m_{H_1}(0)$) the upper limit on $m_{H_2}(0)$ is set by 
demanding the
lightest chargino to be heavier than 90 GeV. For the NMSSM, we have taken
$\lambda(0)=1$ and $k(0)=-0.1$ as a sample point. The values of $A_\lambda(0)$
and $m_N(0)$ are obtained from the minimisation conditions, and we have
restricted the range in $m_{H_2}(0)$ to those values which give rise to a
physical Higgs spectrum not excluded by LEP. For the three values of
$\tan\beta$ shown in the plots, the physical Higgs mass plotted is the second
lightest Higgs, while the lightest one is a dominantly singlet Higgs with a
very weak coupling to the $Z$ boson. In fact, a weakly coupled Higgs as light
as a few GeV might have escaped detection at LEP\footnote{An analytical fit to
the experimental results giving the constraint between the mass of the scalar
Higgs and its coupling to the $Z$ boson can be found in
\cite{seven}.}. 
In both models, MSSM and NMSSM, the increase of the maximum fine-tuning 
when increasing
the physical Higgs mass is mainly due to the increase in $M_3(0)$ \cite{two}.
Varying $m_{H_1}(0)$ and $m_{H_2}(0)$ has almost no effect in fine-tuning in
the MSSM for values of $\tan\beta\geq 3$, and the regions obtained are 
narrower than in the NMSSM.

It is clear that
for all three values of $\tan\beta$ the physical Higgs boson mass in the NMSSM
may be heavier than in the MSSM, with correspondingly lower fine-tuning.
The effect is clearly greater for lower values of $\tan\beta \leq 3$, where
the experimentally allowed values of the lightest Higgs mass would imply large
fine-tuning in the MSSM, whereas in the NMSSM we can have a large enough Higgs
mass with a relatively low $\Delta^{max}$. The plots are a striking 
demonstration that the physical Higgs boson can be heavier and involve
less fine-tuning in the NMSSM compared to the MSSM at low values of
$\tan\beta$. 

\section{Baryogenesis in the NMSSM}

Having shown that the low $\tan\beta$ region of the NMSSM permits a heavier
physical Higgs boson consistent with the LEP limit, without large fine-tuning,
we now turn to the question of electroweak baryogenesis in the NMSSM. The
anomaly mediated electroweak processes are known to provide an efficient
mechanism for baryogenesis in the symmetric phase \cite{three}. In order to
prevent the washout of the resulting baryon asymmetry by the back reaction,
however, one requires a strongly 1st order EWPT,
\beq
v_c > T_c, \label{nine}
\eeq
where $T_c$ is the critical temperature and $v_c$ the Higgs VEV in the 
symmetry
breaking phase at this temperature. This imposes serious constraints on the
underlying model, since it requires a negative cubic term in the generic Higgs
potential
\beq
V = m^2 \phi^2 - \mu \phi^3 + \lambda \phi^4,
\label{ten}
\eeq
with
\beq
v_c = \mu/2\lambda > T_c,
\label{eleven}
\eeq
where the critical point is defined as the point of degeneracy between the
symmetric and symmetry breaking vacua.

In the SM, thermal loops provide a positive quadratic term as well as a
negative cubic term to (\ref{ten}), the latter coming from the $W$ and $Z$
boson loops. Being a loop effect however the resulting $\mu$ is small compared
to $T_c$, which is typically of the electroweak scale. This implies a 
stringent
limit on $\lambda$ and hence on the SM Higgs mass, $m_h \lsim 40$ GeV
\cite{three}, which is ruled out by the LEP data. In the MSSM the cubic term
gets an additional contribution from stop loops; so that for a light stop
$(m_{\tilde t_R} \sim 100~{\rm GeV})$ the Higgs mass limit goes up to $m_h
\lsim 100$ GeV for $m_A \gg M_Z$ \cite{three}. But the twin requirements on
$m_h$ and $m_A$ squeezes the MSSM solution to the low $\tan\beta$
region, 
which
is disfavoured by LEP. Indeed the only way of reconciling the LEP limit with
the low $\tan\beta$ region is by invoking large stop mass and mixing
parameters, in which case one can not satisfy (\ref{nine}). Thus the MSSM
scenario for a strong EWPT (\ref{nine}) is strongly disfavoured by LEP
data, if
not ruled out by it altogether.

As already noted in \cite{eight,nine,ten}, it is much easier to satisfy the
strong EWPT requirement (\ref{nine}) in the NMSSM compared to the MSSM, since
the tree level potential (\ref{four}) itself contains a cubic term $2\lambda
A_\lambda x v_1 v_2$. We closely follow the approach of \cite{eight} in using
the simplest $\ZZ_3$ symmetric form of the NMSSM superpotential 
(\ref{three}) and
simply retaining the leading term in the expansion of the thermal loop in
$M/T$, i.e. the mass of the exchanged particle relative to the temperature.
Thus
\beq
V = V_0 + V_T,
\label{twelve}
\eeq
where $V_0$ is the zero-temperature potential represented by
(\ref{four}) 
along
with the radiative correction from top/stop loops, and
\bea 
V_T & = & {T^2 \over 24} \Big[{\rm Tr}~ M^2_S + {\rm Tr}~M^2_P + 2m^2_{H^+} +
6M^2_W + 3M^2_Z \nonumber \\
& & + 6m^2_t + 2m^2_C + m^2_{N1} + m^2_{N2} + m^2_{N3}\Big]. \label{thirteen}
\eea
The last four terms represent the charged and the neutral Higgsinos. The other
superparticle masses are assumed to be larger than $2T$ and hence 
suppressed by
the Boltzmaan factor \cite{eleven}. We shall evaluate the field-dependent
masses of (\ref{thirteen}) in the Landau gauge as in \cite{nine} 
instead of the
Unitary gauge of ref. \cite{eight}, in view of the well-known ambiguity in
calculating finite temperature effects in the latter (the so-called unitary
gauge puzzle \cite{twelve}). This gives \cite{nine} 
\bea
V_T &=& {T^2 \over 24} \Big[4m^2_{H_1} + 4m^2_{H_2} + 2m^2_N + 2(3g^2 +
g^{\prime 2} + 3\lambda^2) (v^2_1 + v^2_2) \nonumber \\
& & + 6h^2_t v^2_2 + 12(\lambda^2 + k^2)x^2\Big]. \label{fourteen}
\eea
To get the potential as a function of the Higgs fields $H_1,H_2,N$ one has
to simply substitute these quantities for their VEVs $v_1,v_2,x$ in
Eqs.~(4) and (14). 

For a given $\tan\beta$ and stop mass and mixing parameters we first minimize
the $T=0$ potential (\ref{four}) along with the radiative correction
and impose
the experimental constraints on the output Higgs parameters to find an
appropriate set of $\lambda,k,A_\lambda,A_k$ and $x$ parameters. We then find
the $T_c$ as the largest $T$ at which the curvature (second derivative) of the
potential changes sign along any direction at the origin. It is clear from
Eqs.~(\ref{four}), (\ref{five}), (\ref{twelve}) and (\ref{fourteen}) that
\beq
T^2_c = {\rm Max} \left[{-12 m^2_{H_1} \over 3g^2 + g^{\prime 2} 
+ 3\lambda^2},
{-12 (m^2_{H_2} + \Delta_{\rm rad}) \over 3g^2 + g^{\prime 2} + 3\lambda^2 +
3h^2_t}, {-2m^2_N \over \lambda^2 + k^2}\right]. \label{fifteen}
\eeq
where $\Delta_{\rm rad} = {-3h^2_t \over 8\pi^2} m^2_t \ell n {m_{\tilde t}
\over m_t}$ along with a stop mixing contribution. Although the $T_c$
determined from this saddle point of the potential at the origin is not
identical to the one determined by the above mentioned degeneracy between the
two vacua, they have been shown to agree within 2\% \cite{nine}. Therefore we
use this $T_c$ and minimise the potential at this temperature to determine the
corresponding VEVs\footnote{Our minimisation code for the NMSSM potential at
finite termperature is based on the corresponding code of Manuel Drees for the
zero temperature case.}. Since both the doublet Higgs fields contribute to the
sphaleron energy the relevant VEV for the requirement (\ref{nine}) is
\beq
v_c = \sqrt{2} (v^2_{1c} + v^2_{2c})^{1\over2}, \label{sixteen}
\eeq
where the factor $\sqrt{2}$ is due to the normalisation convention, $M^2_Z =
(g^2 + g^{\prime 2}) (v^2_1 + v^2_2)/2$ as emphasised in \cite{thirteen}. 

The stop mass and mixing parameters are fixed at
\beq
m_{\tilde t} = 500~{\rm GeV}, ~A_t = 1000~{\rm GeV},
\label{seventeen}
\eeq
and input values of $\tan\beta$ are taken as $2,3,5$ and $10$. 
\footnote{We obtain similar results for 
$m_{\tilde t} = 250~{\rm GeV}$, $A_t = 500~{\rm GeV}$.}
For each
$\tan\beta$ we run the program at $7 \times 10^5$ points, corresponding to 7
different starting points in the parameter space. Only about 0.1\% of these
points give solutions satisfying the experimental constraints along with the
requirement that in each case the solution represents the absolute minimum of
the potential. This is about an order of magnitude less than the passing rate
of ref. \cite{nine}, which could be due to our using the most constrained
($\ZZ_3$ symmetric) form of the NMSSM as well as the stronger experimental
constraints coming from LEP now. About 10\% of these solutions satisfy the
strong EWPT criterion (\ref{nine}). Thus for each of the 7 starting points in
the parameter space we have $\lsim 10$ solutions satisfying all these 
criteria.

\begin{table}
\[
\begin{tabular}{|c|c|c|c|c|c|c|}
\hline
&&&&&& \\
$\tan\beta$ & $m_{Si}$ & $m_{Pi},m_{H^+}$ & $\lambda,k$ &
$A_\lambda,A_k$ & $x$ & $v_c,T_c$ \\
  & (GeV) & (GeV) & & (GeV) & (GeV) & (GeV) \\
&&&&&& \\
\hline
&&&&&& \\
2 & 82,114,468 & 235,461,464 & .33,-.12 & -340,227 & 621 & 80,76 \\
&&&&&& \\
  & 113,139,664 & 390,659,662 & .31,-.14 & -465,379 & 953 & 162,58 \\
&&&&&& \\
\hline
&&&&&& \\
3 & 97,126,751 & 145,750,747 & .48,-.12 & -658,94 & 485 & 118,81 \\
&&&&&& \\
  & 115,126,975 & 169,977,973 & .41,.10 & -1067,-146 & 704 & 187,61 \\
&&&&&& \\
\hline
&&&&&& \\
5 & 46,127,618 & 84,620,614 & .43,.12 & -690,-107 & 257 & 191,66 \\
&&&&&& \\
  & 122,166,606 & 296,604,609 & .14,-.15 & -510,243 & 798 & 123,82 \\
&&&&&& \\
\hline
&&&&&& \\
10 & 109,127,456 & 146,455,462 & .04,-.10 & -737,102 & 709 & 110,91 \\
&&&&&& \\
   & 120,128,844 & 187,844,847 & .12,.14 & -1082,-145 & 583 & 156,78 \\
&&&&&& \\
\hline
\end{tabular}
\]
\label{table}
\caption{
{\footnotesize Representative sample of NMSSM solutions satisfying all
the experimental constraints along 
with the requirement of a strong EWPT ($v_c
> T_c$). The physical Higgs masses are shown along with the model parameters.
($m_{\tilde t} = 500~{\rm GeV}, A_t = 1000~{\rm GeV}$).}}
\end{table}

Table 1 shows two representative solutions for each $\tan\beta$ along with the
corresponding parameter values as well as the output Higgs boson masses, $v_c$
and $T_c$. One of the solutions corresponds to a relatively low value of the
lightest scalar mass, escaping the LEP constraint because of its large ($\geq
98\%$) singlet content, while the other has a lightest scalar mass above the
LEP limit. We see from this table that it is possible to get acceptable Higgs
mass spectra as well as satisfy $v_c > T_c$ with reasonable values of the
parameters, at least for low values of $\tan\beta$ (2,3,5). Interestingly one
gets acceptable solutions for $\tan\beta = 10$ as well; but in this case $k >
\lambda$. The reason is that for large $\tan\beta$ 
the 3rd and 4th terms of the
potential (\ref{four}) become vanishingly small. The latter implies that the
$m^2_h$ limit of Eq.~(\ref{two}) is dominated by the 1st term, 
which is however
quite large now; while the former implies that the dominant cubic term of the
potential is ${2\over3} k A_k x^3$. Thus the NMSSM can help to get a strong
EWPT in the large $\tan\beta$ region as well, 
where it has little effect on the
Higgs mass limit of the MSSM.

It is appropriate to briefly discuss here the role of NMSSM in generating
sufficient amount of CP asymmetry, as required for baryogenesis
\cite{thirteen,fourteen}. 
Of course quantitative investigation of this question
depends on the model of electroweak baryogenesis, which is beyond the scope of
this work. But it is generally agreed that the size of CP violation in the SM,
arising from the complex CKM matrix, is much too small as it is suppressed by
the small Yukawa couplings as well as the CKM mixing angles. There are
additional sources of CP violation in the MSSM, arising from the
phases of $\mu$ and 
the SUSY breaking terms, which can serve this purpose provided the size of the
phase angles are $\geq O(10^{-1})$ \cite{three}. On the other hand the
experimental constraint from the electric dipole moments of neutron and
electron would require these phase angles to be $\leq O(10^{-2})$ unless there
is a systematic cancellation between them \cite{fifteen} or one assumes the
sfermions of the 1st two generations to have 
masses $\gg 1$ TeV \cite{sixteen}.
This potential conflict with the electric dipole moment 
limits is alleviated in
the NMSSM, where the required size of phase angles is an order of magnitude
smaller than in MSSM \cite{thirteen}. Even more interestingly the NMSSM offers
the possibility of generating a spontaneous CP violation in the symmetric
phase, which goes down to zero in the symmetry breaking phase
\cite{thirteen,fourteen}. Thus it can effectively contribute to baryogenesis,
which takes place in the symmetric phase, while making no contribution to the
measured electric dipole moments. It is not possible to generate such a
transitional CP violation in the MSSM \cite{thirteen}.

\section{Conclusion}

The current LEP limit on the lightest Higgs boson mass places severe
constraints on the MSSM in the low $\tan\beta$ region, which was the favoured
region for a strong EWPT as required for electroweak baryogenesis. 
The only way
to escape this $m_h$ limit is to invoke very large stop mass and mixing
parameters, which would imply however large fine-tuning as well 
as a weak EWPT.
Thus one has to sacrifice the naturalness of electroweak symmetry breaking as
well as baryogenesis. However both these problems can be solved simultaneously
with the so called $\mu$ problem of the MSSM by going to the NMSSM. All the
relevant terms -- i.e. $\lambda^2 x^2 v^2_{1(2)}$, $\lambda^2 v^2_1 v^2_2$ and
$\lambda A_\lambda x v_1 v_2$ -- originate from the superpotential $\lambda N
H_1 H_2$. While the 1st term solves the so-called $\mu$ problem, the 2nd
provides an additional contribution to the tree-level $m_h$ limit and the 3rd
one a cubic term in the tree-level potential. Thus the 2nd term alleviates the
fine-tuning problem arising from the $m_h$ limit, while the 3rd ensures a
strong EWPT.

It may be noted here that both the 2nd and the 3rd terms vanish at large
$\tan\beta$. Thus it does not affect the $m_h$ limit of the MSSM at large
$\tan\beta$, which is any way quite high. However in this case the soft cubic
term $k A_k x^3$ helps to generate a strong EWPT. Thus the NMSSM helps to give
a strong EWPT along with solving the $\mu$ problem even in the large
$\tan\beta$ region.

To summarise, the LEP limit on the Higgs boson mass severely 
constrains the low
$\tan \beta$ region of the MSSM, leading to large fine-tuning and 
problems with
electroweak baryogenesis. We have shown that in the low $\tan \beta$ region
the NMSSM is in much better shape
phenomenologically, since the physical Higgs boson masses are larger, the
fine-tuning is less, and the electroweak phase transition is more strongly
first order in the NMSSM, as compared to the MSSM.

Finally we remark that although we have considered the NMSSM for
simplicity, we would expect similar effects in more complicated
extensions of the NMSSM, for example those 
involving an additional anomalous $U(1)$
gauge group \cite{lisa}, or those including Higgs triplets \cite{jose},
since both models also involve the $\lambda N H_1 H_2$ coupling
considered here.

\begin{center}{\bf Acknowledgements}\end{center} The authors would like to
thank the organisers of WHEPP-6, where this project was started, and B.
Ananthanarayan and P.N. Pandita for their contributions during the initial
stage of the work. We would also like to thank C.D. Froggatt and M. Pietroni
for several helpful communications.

\end{document}